\documentclass[reprint,prl,showpacs,superscriptaddress]{revtex4-2}
\usepackage{amssymb}
\usepackage{amsmath}    
\usepackage{graphicx}   
\usepackage{verbatim}

\usepackage{color}      
\usepackage{booktabs} 

\usepackage[hidelinks]{hyperref}

\begin{document}

\title{Reaching the high laser intensity by a radiating electron}

\author{M. Jirka}
\affiliation{ELI Beamlines Centre, Institute of Physics, Czech Academy of 
Sciences, Za Radnici 835, 25241 Dolni Brezany, Czech Republic}
\affiliation{Faculty of Nuclear Sciences and Physical Engineering, Czech
Technical University in Prague, Brehova 7, 115 19 Prague, Czech Republic}

\author{P. Sasorov}
\affiliation{ELI Beamlines Centre, Institute of Physics, Czech Academy of 
Sciences, Za Radnici 835, 25241 Dolni Brezany, Czech Republic}
\affiliation{Keldysh Institute of Applied Mathematics, Moscow, 125047, Russia}

\author{S. S. Bulanov}
\affiliation{Lawrence Berkeley National Laboratory, Berkeley, California 94720, 
USA}

\author{G. Korn}
\affiliation{ELI Beamlines Centre, Institute of Physics, Czech Academy of 
Sciences, Za Radnici 835, 25241 Dolni Brezany, Czech Republic}

\author{B. Rus}
\affiliation{ELI Beamlines Centre, Institute of Physics, Czech Academy of 
Sciences, Za Radnici 835, 25241 Dolni Brezany, Czech Republic}

\author{S. V. Bulanov}
\affiliation{ELI Beamlines Centre, Institute of Physics, Czech Academy of 
Sciences, Za Radnici 835, 25241 Dolni Brezany, Czech Republic}
\affiliation{National Institutes for Quantum and Radiological Science and 
Technology (QST), Kansai Photon Science Institute, 8-1-7 Umemidai, Kizugawa, 
Kyoto 619–0215, Japan}

\begin{abstract}
Interaction of an electron with the counter-propagating electromagnetic wave is 
studied 
theoretically and with the particle-in-cell simulations in the regime of 
quantum radiation reaction.
We find the electron energy in the center of the laser pulse, as it is a key 
factor for testing the non-linear quantum electrodynamics vacuum properties in 
the laser-electron 
collision in the regime of multi-photon Compton scattering and vacuum Cherenkov 
radiation.
With multiparametric analysis we provide the conditions on electron 
initial energy for reaching the center of the laser pulse and emitting 
Cherenkov photons.
\end{abstract}

\maketitle
One of the interaction geometries employed to study the strong field Quantum Electrodynamics (SF QED) is the collision of a high energy electron beam and an intense electromagnetic (EM) pulse \cite{Bulanov2011,Thomas2012,DiPiazza2012,Blackburn2020,Bamber1999,Cole2018,Poder2018}. With the increase of laser intensities \cite{Mourou2006,Danson2019,ELIwhitebook,Weber2017,Tanaka2020} as well as the energies of laser-accelerated electrons \cite{Gonsalves2019}, the utilization of them both turned in to a standard setup for the study of radiation reaction (RR), both classical and quantum, which starts to dominate electron energy loss through photon emission (Compton process). The photons being exposed to the strong EM fields right after the emission can transform into electron-positron pairs through the Breit-Wheeler process \cite{Bell2008,Thomas2012}. Since the rate of this transformation depends on both photon energy and field strength, which in their turn depend on how far an electron penetrated into the region of strong fields in the laser pulse, the understanding of electron energy depletion due to photon emission is of paramount importance. In other words, the question whether a high energy electron can reach the strong field region in the presence of the RR and how its energy depends on the laser pulse phase needs to be answered \cite{Bulanov2013,Blackburn2014,Vranic2014,Blackburn2017,Seipt2017}.


When laser intensities reach above $ 10^{23}~\mathrm{W/cm^{2}} $ for $1~\mathrm{\mu m} $ wavelength, the QED effects in the laser-electron collision can significantly alter the electron dynamics affecting the energy spread and the divergence of the electron beam \cite{Neitz2013,Green2014,Vranic2016Quantum,Ridgers2017}. Further increase of both laser intensity and electron beam energy will lead to a number of exotic effects to become observable, such as Cherenkov radiation \cite{Ritus1970,Ritus1985,Dremin2002,Macleod2019,Bulanov2019,Artemenko2020}, vacuum birefringence \cite{Bragin2017}, and the regime where the semi-perturbative expansion of the SF QED supposedly breaks down \cite{Yakimenko2019,Blackburn2019,Zhang2020}. 


The strong fields mentioned above are those that either approach or exceed the critical field of QED,  $E_{\mathrm{S}} =m_{e}^{2}c^{3}/e\hbar\approx1.33\times10^{16}~\mathrm{V/cm} $, where $ m_{e} $ denotes the electron mass, $ c $ speed of light, $ e $ is the elementary charge and $ \hbar $ is the reduced Planck constant. It corresponds to the electric field strength needed for the creation of electron-positron pairs out of the vacuum. However, to achieve such a field strength with a laser, one needs the power of $ \mathcal{P}\sim\left( \lambda\left[ \mu m\right] \right)^{2} ~\mathrm{ZW} $ provided that the laser pulse is focused to a $ \lambda $ spot-size. The corresponding energy of a single cycle laser is $ \mathcal{E}\sim 3\left( \lambda\left[ \mu m\right] \right)^{3} ~\mathrm{MJ}$. 
Therefore the reduction of the laser wavelength will result in lower power and energy requirements to achieve high intensity. 
Generation of ultra-short pulses with a required intensity is outside performance of the existing lasers. 
However, several possible paths to generate ultra-short laser pulses focusable to intensities $ 10^{25}~\mathrm{W/cm^{2}} $ have been proposed or are conceptually possible by extrapolating the existing laser technology. 
These include using self-phase modulation of a broadband multi-petawatt laser pulses in a structured solid medium \cite{Voronin2013}, prospects for generation of ultra-high contrast second harmonic fs pulses at 400 nm in a Ti:sapphire system \cite{Wang2018} or at $527~\mathrm{nm}$ from high-energy Nd:glass kJ beams \cite{Hopps2015} by ultrathin frequency converters, prospects for generation of a $ \approx530~\mathrm{nm}$, $<20~\mathrm{fs} $ pulses by OPCPA in a large-aperture LBO amplifier driven by third harmonics of a $ 1~\mathrm{\mu m} $ pump laser \cite{Endo16,Antipenkov2020}.
Another technique for shortening the laser wavelength is to exploit the non-linear properties of the laser-produced plasma. This is represented by the concept of relativistic flying plasma mirrors  \cite{Bulanov2003,Bulanov2013RelativisticMirrors}.
%
%
%
Thus, it is important for future experiments to reveal how the impact of QED processes on the high-intensity laser-matter interaction scales with the laser wavelength, energy and intensity \cite{Zhang2020}.

%
%
In this letter we consider the interaction of the 10--100s~GeV electron with an EM wave of intensity $ 10^{23-25}~\mathrm{W/cm^{2}} $.  Using geometrical correspondence between the flat-top and Gaussian temporal envelope  of the laser pulse delivering the same energy we provide straightforward estimates of the electron energy in the center of the laser pulse and of the threshold for electron reflection considering the quantum regime of RR. These factors are of great importance for future experiments requiring the electron of specific energy experiencing the amplitude of the laser field. Then we utilize these estimates for finding the laser parameters allowing us to observe the Cherenkov radiation in the laser-electron collision providing the signature of vacuum polarization in the distribution of Breit-Wheeler positrons.

In QED the interaction of charged particles (electrons and positrons) and photons with EM field is characterized by two Lorentz invariant parameters, $ \chi_e=\sqrt{-\left( F^{\mu\nu}p_{\nu}\right)^{2} }/m_{e}cE_{\mathrm{S}} $ and $ \chi_\gamma= \hbar\sqrt{-\left( F^{\mu\nu}k_{\nu}\right)^{2}}/m_{e}cE_{\mathrm{S}}$, where $ F_{\mu\nu}=\partial_{\mu}A_{\nu}-\partial_{\nu}A_{\mu}$ is the EM field tensor, $ A_{\nu} $ is the four-potential, $ p_{\nu} $ is the four-momentum of the electron, and $ k_{\nu} $ is the photon four-wave vector. The strength of the EM field is characterized by the Lorentz invariant parameter $a_{0}=eE_{0}/m_e\omega_{0} c $, where $ E_{0} $ is the amplitude of the electric field and $ \omega_{0} $ is the laser frequency. We assume a head-on collision of an electron having the initial energy $ \mathcal{E}_{e}=\gamma_{e} m_{e}c^{2} $, where $ \gamma_{e}\gg1 $ is the relativistic Lorentz factor, with a linearly polarized plane wave characterized by the peak intensity $ I_{0} $, the wavelength $ \lambda $ and the full width at half maximum duration $ \tau $ in laser intensity. In this case, the maximum value of parameter $ \chi_{e} $ can be approximately expressed as $ 2\gamma_{e}E_{0}/E_{\mathrm{S}} $.

In the limit of a large $ \chi_{e} $ parameter ($ \chi_{e}\gg 1 $), i.e. in the quantum RR regime, the probability of single-photon emission per unit time for the Compton process is given by
%
$ W_{\gamma}\approx 3^{2/3}28\Gamma\left( 2/3\right)\alpha 
m_{e}^2c^{4}\chi_{e}^{2/3}/54\hbar\mathcal{E}_{e},$
%
where $ \Gamma\left( x\right)  $ is the Gamma function and $ \alpha=e^{2}/\hbar c $ is the fine structure constant. As a result of a photon emission an electron losses $16/63$ of its initial energy on average \cite{Ritus1985}.

In order to estimate an electron energy loss, we assume that the electron travels towards the high-intensity region of the laser pulse. The electron emits Compton photons according to the probability $ W_{\gamma}$. Each emission leads to the above-mentioned electron energy loss.
Further we assume that the electron always emits one photon when $ W_{\gamma}\Delta t = 1 $, where $ \Delta t $ is the time of radiation. To adapt $ W_{\gamma} $ for a laser pulse with a Gaussian temporal envelope characterized by the peak intensity $ I_{0} $ and the duration $ \tau $, we need to consider a laser pulse with a flat-top temporal envelope having the duration $ 2\tau/\sqrt{2\ln2} $ providing the same energy as the Gaussian one. The time required for an electron to achieve the center of the top-hat laser pulse is $ \tau/\sqrt{2\ln2} $.

Photon emission probability depends on the phase $ \psi=\omega_{0}t $ of the laser field as $ \sin\psi $. For a single laser cycle, the probability is the highest twice per a laser period $ T $ when the intensity $ I\propto\sin^{2}\psi $ reaches its local maxima at $ \psi=\pi/2 $ and $ \psi=3\pi/2 $.
Therefore, we assume that photon emission can happen when $ \sin^{2}\psi\ge1/2 
$
%
%
%
and thus the effective time interval for photon emission by the moment when the electron reaches the center of the laser pulse is $ t_{\mathrm{c}}\approx \tau/2\sqrt{2\ln2} $. During the time interval $ t_{\mathrm{c}} $, multiple emission of photons may occur.
Since the number of emitted photons  can be estimated as $ p_{\mathrm{c}}\approx W_{\gamma}t_{\mathrm{c}} $, the electron reaches the center of the laser pulse with the energy given by
\begin{equation}\label{eps_c}
\mathcal{E}_{e}^{\mathrm{c}} \approx \left( 1- 
16/63\right)^{p_{\mathrm{c}}}  \mathcal{E}_{e}.
\end{equation}
Then the average $ \chi_{e}^{\mathrm{c}} $ of the electron in the presence of the laser field amplitude can be directly obtained as $ \chi_{e}^{\mathrm{c}}\approx 2\gamma_{e}^{\mathrm{c}}E_{0}/E_{\mathrm{S}} $, where $ \gamma_{e}^{\mathrm{c}} $ is the relativistic Lorentz factor of the electron in the center of the laser pulse. 
The final electron energy after passing through 
the laser pulse is
$ \mathcal{E}_{e}^{\mathrm{f}} \approx \left( 1- 
16/63\right)^{p_{\mathrm{f}}}   \mathcal{E}_{e} $,
where $ p_{\mathrm{f}}\approx W_{\gamma}t_{\mathrm{f}} $
and 
$ 
t_{\mathrm{f}}\approx \tau/\sqrt{2\ln2}.
 $

We assume that the electron has initially enough energy to reach the center of the laser pulse. However, even when RR is not considered, the electron can be prevented from experiencing the laser field amplitude due to the ponderomotive force if $ \gamma_{e} < \sqrt{1+a_{0}^{2}/2} $. When RR comes into play, then the electron with $ \gamma_{e}\gg \sqrt{1+a_{0}^{2}/2} $ can lose a significant fraction of its energy as it propagates towards the laser pulse center. The condition for electron reflection therefore depends on the ponderomotive potential barrier and the actual energy of the radiating electron. As a result, the electron can be expelled by the ponderomotive force before reaching the  center of the laser pulse. In the limit $ \chi_{e}\gg1 $, we estimate, using the above-presented approach, the threshold for reflection of the radiating electron as
\begin{equation}\label{reflection}
\mathcal{E}_{e}^{\mathrm{r}}/m_{e}c^{2} 
\approx\sqrt{1+a_{0}^{2}/2} ,
\end{equation}
where $ \mathcal{E}_{e}^{\mathrm{r}} \approx \left(1-16/63 \right)^{p_{\mathrm{r}}}\mathcal{E}_{e} $, $ p_{\mathrm{r}}\approx W_{\gamma}t_{\mathrm{r}} $ and $ t_{\mathrm{r}}\approx t_{\mathrm{c}}/\sqrt{2}\left( \sqrt{2\ln2}\right)^{\tau/T -1} $ that accounts for the conversion between the Gaussian and the flat-top laser pulse. We note, however, that the consideration of the electron reflection can not be achieved in the framework of an adopted 1D model. In the case when the condition (\ref{reflection}) becomes relevant, the transverse momentum of electron becomes of the order of longitudinal one and full 3D dynamics of the electron should be taken into account. 

Having written down the estimates for the average electron energies in the center of the pulse and after the interaction, we compare them with the results of 1D Particle-In-Cell (PIC) simulations performed by the code {\sf SMILEI} \cite{Derouillat2018}. In the code, photon emission and electron-positron pair creation are modelled using Monte-Carlo approach. We use a linearly polarized plane wave with a Gaussian temporal envelope $ \tau $ having the peak intensity $ I_{0} $ in the range $ 10^{23-25}~\mathrm{W/cm^{2}} $ and the wavelengths $ \lambda=0.25~\mathrm{\mu m} $, $ 0.5~\mathrm{\mu m} $ and $ 1~\mathrm{\mu m} $. Such a laser pulse collides head-on with $ 10^{5} $ simulation electrons of initial energy $ \mathcal{E}_{e} $.

\begin{figure}
	\centering
	\includegraphics[width=1.0\linewidth]{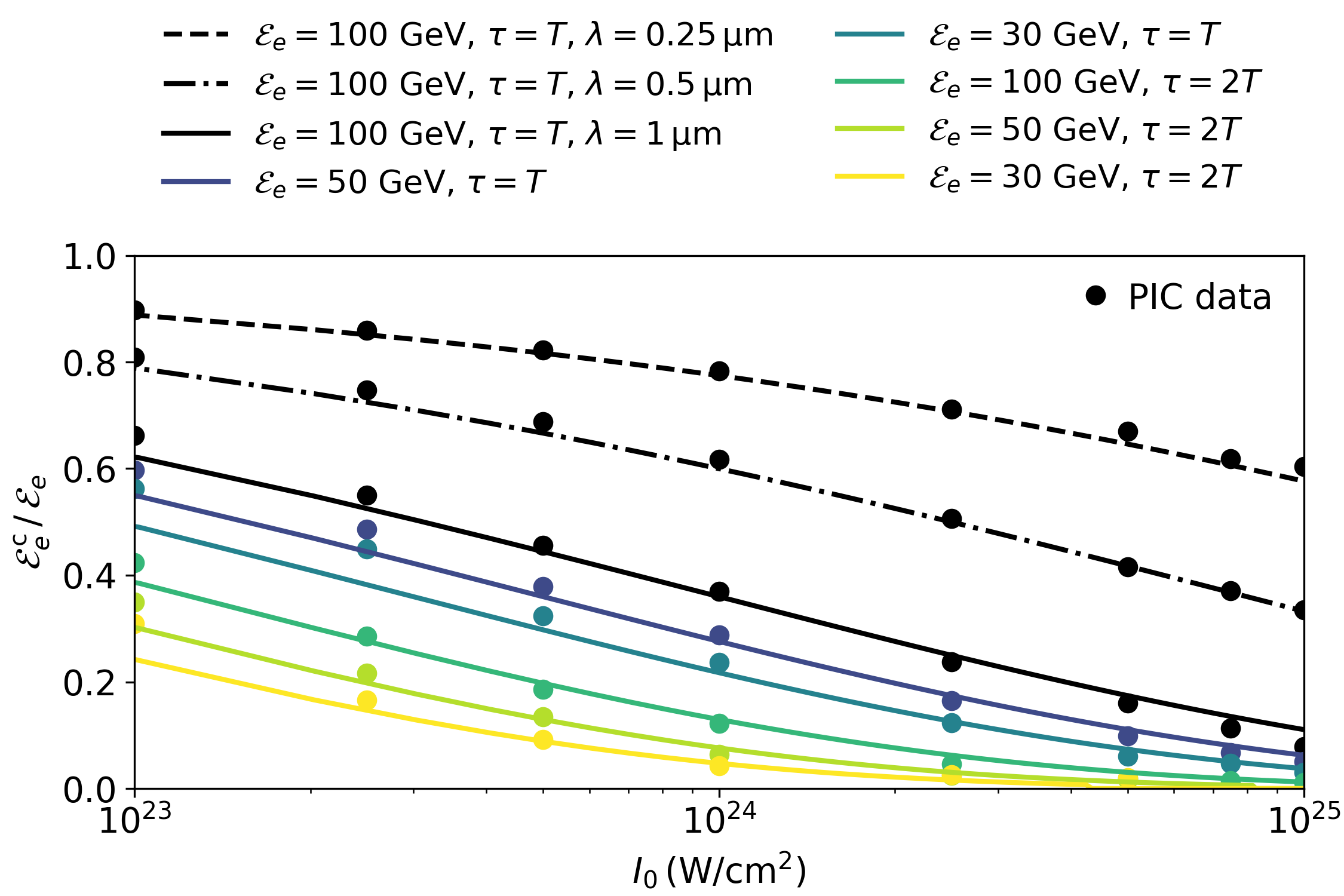}
	\caption{Electron energy $ 
			\mathcal{E}_{e}^{\mathrm{c}} $ in the center of the laser pulse as 
			a function of laser intensity $ I_{0} $ predicted 
		by 
		Eq.~\eqref{eps_c} (lines) and obtained from PIC 
		simulations (bullets). The laser pulse characterized by the duration $ 
		\tau $ and the wavelength $ \lambda $ collides 
		head-on with electrons of the initial energy $ \mathcal{E}_{e}$. Solid 
		lines represent cases with $ \lambda=1~\mathrm{\mu m} $.}
	\label{fig:Fig_01}
\end{figure}

In Fig.~\ref{fig:Fig_01} we present the results for electron energy in the center of the laser pulse. The theoretical predictions are compared with simulation data for two different durations of the laser pulse and three values of the initial electron energy (solid lines).
%
%
It is shown, that increasing the initial electron energy or shortening the laser pulse duration results in a reduction of the energy emitted before the electron enters the center of the laser pulse. The first aspect is given by the fact that in the limit $ \chi_{e}\gg1 $, the probability of photon emission $ W_{\gamma}\propto\gamma_{e}^{-1/3} $. The latter is caused by a linear dependence of the emitted photon number on the interaction duration.
%
%
Dash-dotted and dashed lines show that shortening the laser wavelength can further reduce the amount of emitted energy before the electron reaches the center of a single cycle laser pulse. 
%
%

\begin{figure}
	\centering
	\includegraphics[width=1.0\linewidth]{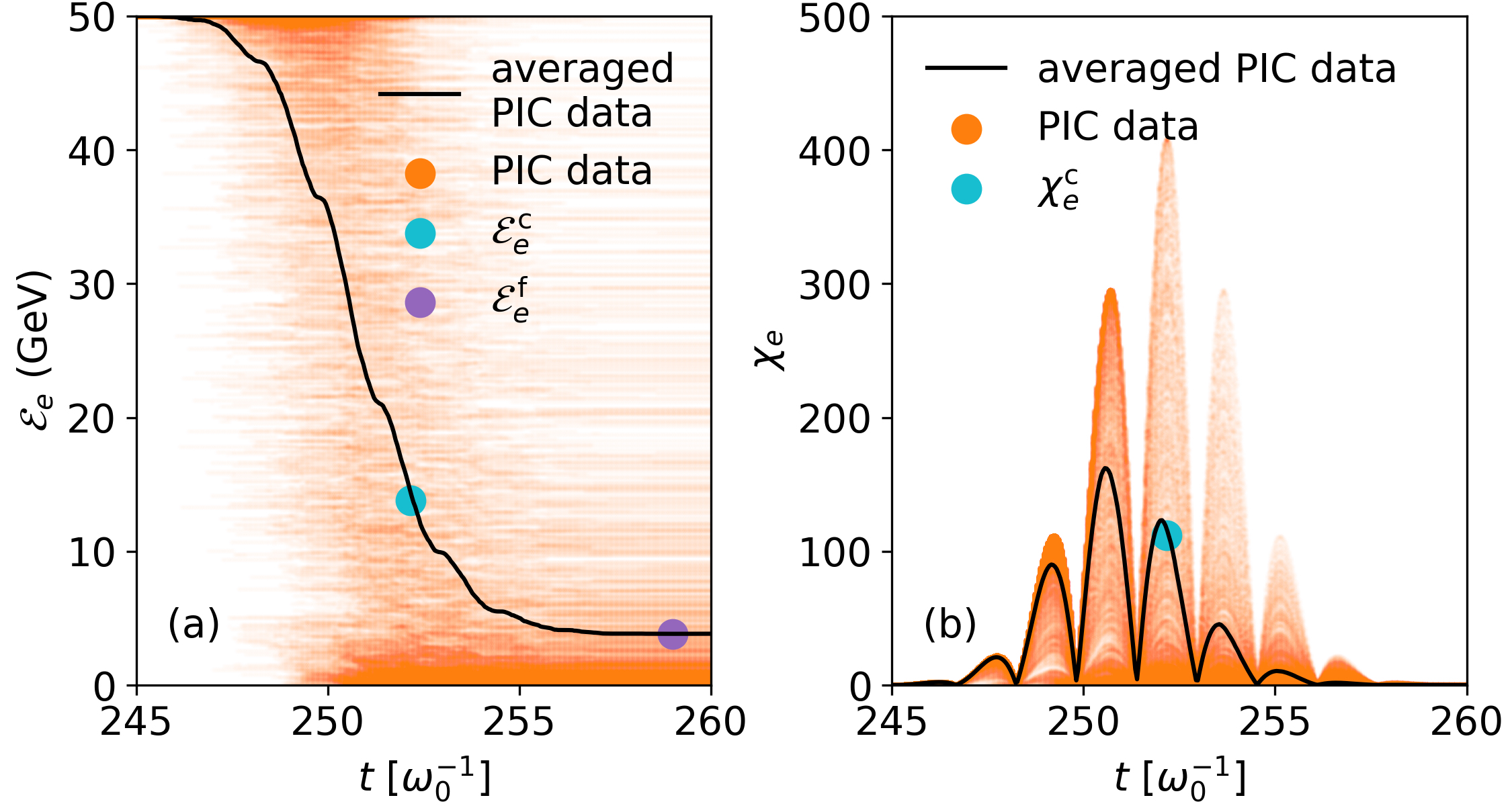}
	\caption{Comparison of PIC data (orange) and their averaged value (black 
		line) with theoretical estimates of electron energy (a) in the center 
		of the laser pulse $ 
		\mathcal{E}_{e}^{\mathrm{c}} $ and after passing the laser pulse $ 
		\mathcal{E}_{e}^{\mathrm{f}} $. (b) Time evolution of $ \chi_{e} $ and 
		the 
		estimated value $ \chi_{e}^{\mathrm{c}} $ in the center of the laser 
		pulse. The laser pulse 
		characterized by the intensity $ I_{0}=10^{24}~\mathrm{W/cm^{2}} $, 
		the wavelength $		
		\lambda=1~\mathrm{\mu m} $ and the duration $ 
		\tau=T $ collides 
		head-on with an electron of the initial energy $ 
		\mathcal{E}_{e}=50~\mathrm{GeV} $.}
	\label{fig:Fig_02}
\end{figure}

Due to the stochastic nature of photon emission in the QED regime of 
interaction, the colliding electrons lose a various fraction of their energy at 
different times.
This can be seen in Fig.~\ref{fig:Fig_02}(a) showing the simulation data of 
electron energy distribution
as a 
function of time for the electron of the initial energy $ 
\mathcal{E}_{e}=50~\mathrm{GeV} $ 
interacting with a laser pulse characterized by $ 
I=10^{24}~\mathrm{W/cm^{2}} $, $ \lambda=1~\mathrm{\mu m} $ and $ \tau=T $.
However, the general trend of the electron energy evolution is illustrated by 
the 
black 
solid line obtained by the averaging of the PIC results.
Its value in the center of the laser pulse and after the 
interaction is in good agreement with theoretical estimations $ 
\mathcal{E}_{e}^{\mathrm{c}} $ and $ \mathcal{E}_{e}^{\mathrm{f}} $, 
respectively.
In panel (b) we show the distribution of parameter $ \chi_{e} $ 
during 
the 
interaction as obtained from PIC simulation.
As can be seen, its average value (black line) is 
considerably affected by RR making it not 
symmetric around its peak value \cite{Bulanov2011}.
However, the value in the center of the laser pulse corresponds with the 
predicted $ \chi_{e}^{\mathrm{c}} $.

\begin{figure}
	\centering
	\includegraphics[width=1.0\linewidth]{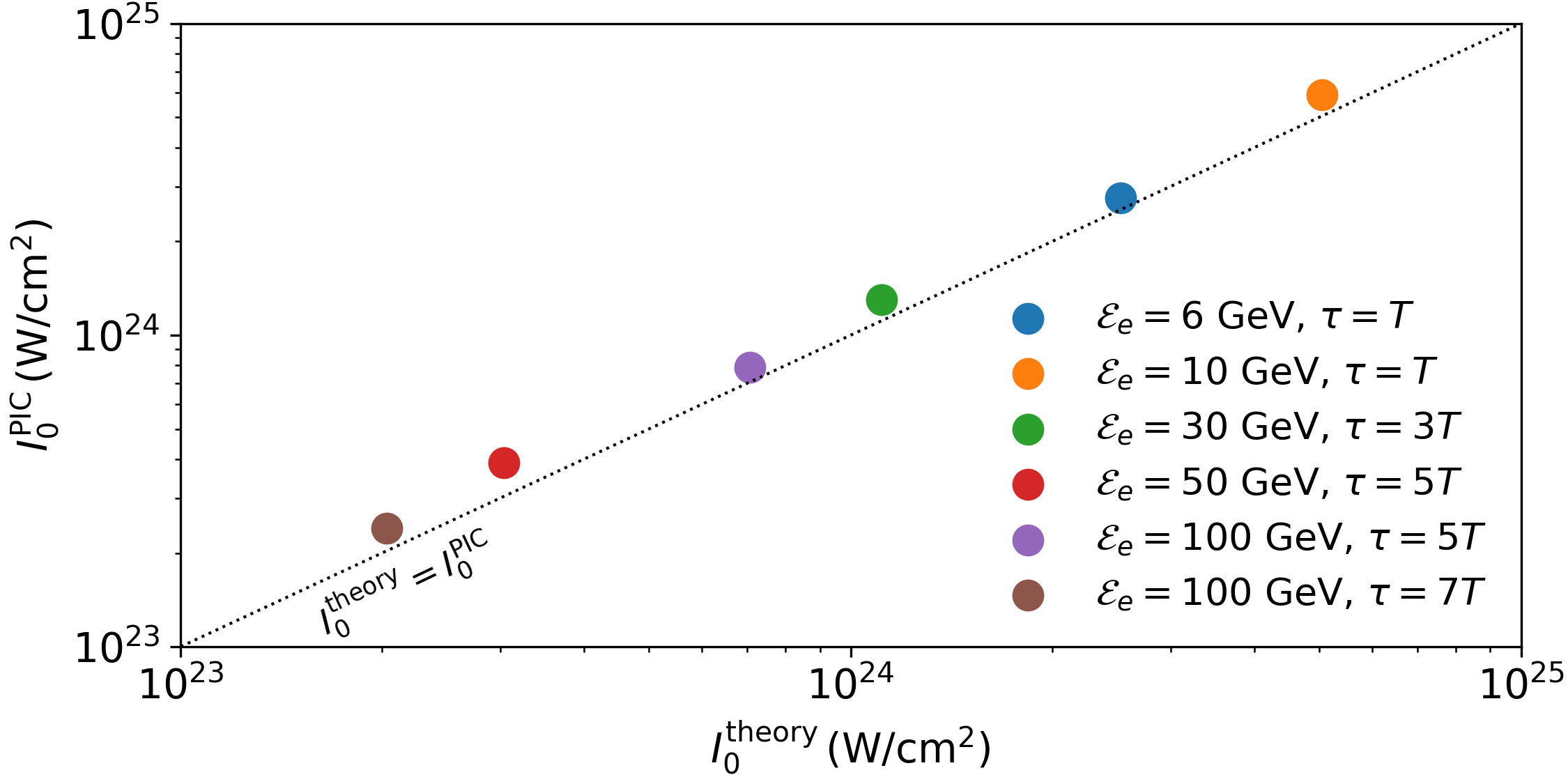}
	\caption{Threshold intensity required for electron reflection by the laser 
	pulse predicted by the theory $I_{0}^{\mathrm{theory}} $ given by 
	Eq.~\eqref{reflection} ($ x $-axis), and 
		obtained from 
		PIC simulations $ 
		I_{0}^{\mathrm{PIC}} $ (bullets) for different values of the initial 
		electron 
		energy $ 
		\mathcal{E}_{e} $ and the laser pulse duration $ \tau $. The laser 
		wavelength is $ \lambda=1~\mathrm{\mu m} $. The dotted line is added to 
		guide the eye for $I_{0}^{\mathrm{theory}} =I_{0}^{\mathrm{PIC}}$.}
	\label{fig:Fig_03}
\end{figure}

In the laser-electron collision, only electrons of sufficiently high energy 
can overcome the ponderomotive barrier created by the amplitude of the intense 
laser pulse.
Using PIC code, we have performed a set of simulations for different 
combinations of 
$ \mathcal{E}_{e} $ and $ \tau $ to find the threshold 
intensity $ I_{0}^{\mathrm{PIC}} $ at which the electrons are reflected.
As shown in Fig.~\ref{fig:Fig_03}, the obtained data are in good agreement 
with 
the threshold intensity $ 
I_{0}^{\mathrm{theory}} $ estimated by Eq.~\eqref{reflection}.
%
%
%
%

The presented approach may help to find the optimal condition for laser-matter 
interaction in which the electron of the required energy is assumed to 
experience 
the amplitude of the laser field.
One example is the emission of the Cherenkov photons in the laser-electron 
collision  
\cite{Ritus1970,Ritus1985,Dremin2002,Macleod2019,Bulanov2019,Artemenko2020}.
The strong EM field of the laser pulse changes the vacuum index of 
refraction, and thus the colliding electron propagating with the 
super-luminal phase velocity can emit the Cherenkov photons.
For observing the Cherenkov radiation in such a collision, the minimum required 
Lorentz factor of the electron experiencing the laser pulse amplitude is given 
by 
$ \gamma_{\mathrm{min}}=1/\sqrt{2\Delta n} $,
where
$ \Delta n = n - 1 $
is the induced change in the index of refraction for a probe photon having $ 
0<\chi_{\gamma}\lesssim15 $ \cite{Ritus1970,Dremin2002}.
%
While $ \Delta n $ has a maximum at $ \chi_{\gamma}\approx5 $, for $ 
\chi_{\gamma}\gtrsim15 $ the value of $ \Delta n  $ becomes negative and thus the
Cherenkov 
radiation vanishes \cite{Ritus1970,Dremin2002,Bulanov2019}.
%
%
In the limit $ \chi_{\gamma} \ll 1$, $ \Delta n =8\alpha E_{0}^{2}/
45\pi E_{\mathrm{S}}^{2} $ \cite{Ritus1970}.

\begin{figure}
	\centering
	\includegraphics[width=1.0\linewidth]{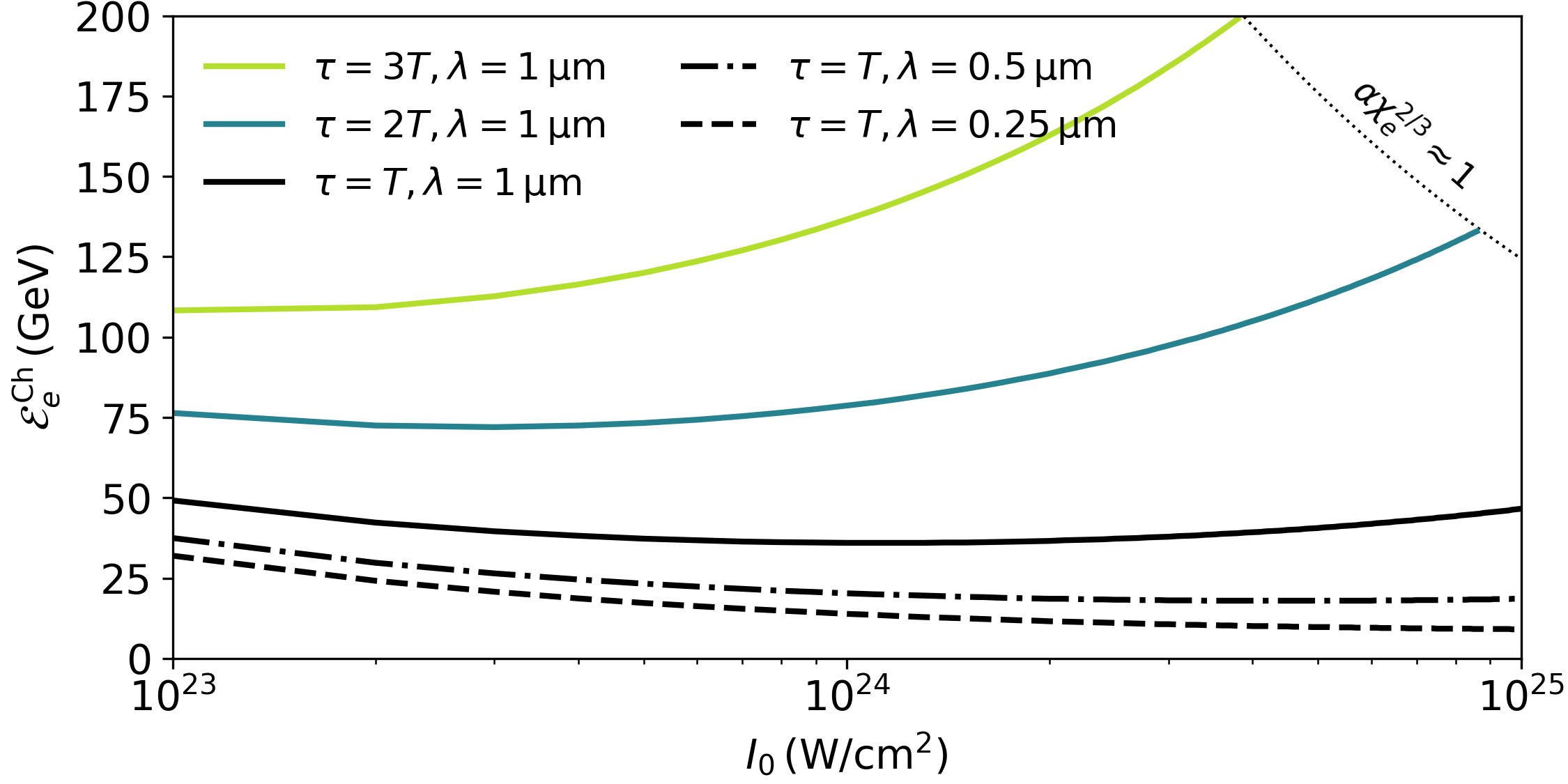}
	\caption{The minimum required initial electron energy $ 
		\mathcal{E}_{e}^{\mathrm{Ch}} $ 
		for the emission of the Cherenkov radiation in the center of the laser 
		pulse of 
		intensity $ I_{0} $, duration $ \tau $ and wavelength $ \lambda $ 
		obtained by Eq.~\eqref{eps_c} for 
		 $ \gamma_{e}^{\mathrm{c}}=		\gamma_{\mathrm{min}}$ in the limit $ 
		 \chi_{\gamma}\ll1 $. }
	\label{fig:Fig_04}
\end{figure}

The Cherenkov photons can only be emitted if the electron in the 
center of the laser pulse satisfies  $ \gamma_{e}^{\mathrm{c}}\ge 
\gamma_{\mathrm{min}}$.
By inserting 	 $ \gamma_{e}^{\mathrm{c}}=
\gamma_{\mathrm{min}}$
into Eq.~\eqref{eps_c},
we 
can estimate what would 
be then the required
minimum initial energy $ \mathcal{E}_{e}^{\mathrm{Ch}} $ of such an electron.
In Fig.~\ref{fig:Fig_04} we present the required initial energy of the electron 
for the emission of the Cherenkov radiation in the center of the laser pulse 
for $ 
\chi_{\gamma}\ll1 $.
It is shown, that using both the short wavelength and the short 
laser duration efficiently reduces the energy threshold.
For example, to achieve the Cherenkov radiation at near-future laser system 
characterized by intensity $ I_{0}=10^{24}~\mathrm{W/cm^{2}} $ and 
pulse 
duration $ \tau=T $, the electron beam of initial energy $ 
\mathcal{E}_{e}^{\mathrm{Ch}}\approx14~\mathrm{GeV} $ is required 
assuming 
the 
wavelength $ \lambda =0.25~\mathrm{\mu m} $.
This provides more than two times lower requirement on the initial electron energy 
compared to the case of $ 1~\mathrm{\mu m} $ laser wavelength.
The presented calculations were performed within the regime where the 
perturbation theory can be applied, i.e. where $ \alpha \chi_{e}^{2/3}\leq1 $.

\begin{figure}
\centering
\includegraphics[width=1.0\linewidth]{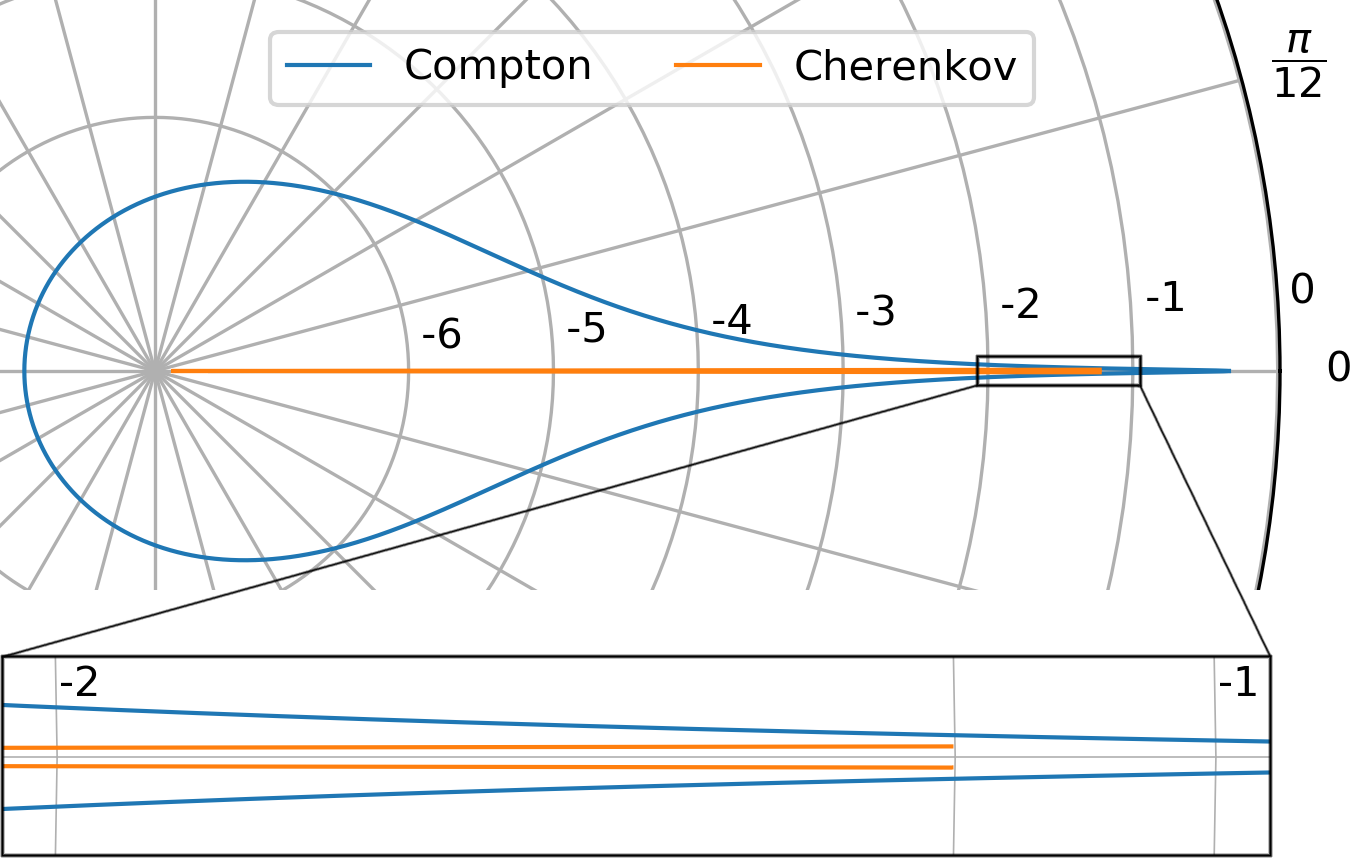}
\caption{The angular energy distribution $  \log_{10}\left[ 
E_{\gamma}\left(\mathrm{GeV} 
\right)\right]  $ of 
Compton photons (blue) emitted by the electron with the initial energy 
$\mathcal{E}_{e}^{\mathrm{Ch}}= 
20~\mathrm{GeV} $ in 
the center of the laser pulse of parameters $I_{0}= 
10^{25}~\mathrm{W/cm^{2}} $, $ \lambda=1~\mathrm{\mu m} $, $ \tau=T 
$. 
Cherenkov photons characterized by $ \chi_{\gamma}\approx1 $ are emitted by 
this electron
within the opening angle $ \theta_{\mathrm{Ch}}\approx 
3\times10^{-3}$ (orange).}
\label{fig:Fig_05}
\end{figure}

%
The presence of the Cherenkov photons in the laser-electron collision can be indicated by the creation of Breit-Wheeler positrons provided that the photon energy is sufficiently high. If the Cherenkov radiation is present, the total number of Breit-Wheeler positrons should be higher than expected from solely Compton radiation. Therefore, we consider the Cherenkov radiation for $ \chi_{\gamma}\approx 1 $ as this represents an optimal regime for SF QED effects  \cite{Ritus1970,Ritus1985,Bulanov2011}. Using the above-mentioned approach, one can identify the lowest electron initial energy for a given laser intensity and duration in order to reach $ \chi_{\gamma}\approx 1 $ for the Cherenkov photons within the limit of perturbative QED. For $ \chi_{\gamma}\approx 1 $ we obtain $ \Delta n \approx 2\alpha E_{0}^{2}/E_{\mathrm{S}}^{2} $ \cite{Ritus1985},
%
and, thus, for example, the required minimum initial electron energy for emitting the Cherenkov photons characterized by $ \chi_{\gamma}\approx 1 $ is $\mathcal{E}_{e}^{\mathrm{Ch}}\approx 20~\mathrm{GeV} $ considering  $I_{0}=10^{25}~\mathrm{W/cm^{2}} $, $ \lambda=1~\mathrm{\mu m} $ and  $ \tau=T $. Such a photon carries out $\chi_{\gamma}/\chi_{e}\approx1/12  $ of the electron energy when radiated in the center of the laser pulse. One can compare the importance of the Cherenkov and the Compton radiation of such photons. The emitted power by a photon with  $ \chi_{\gamma}\approx1 $ is $P_{\mathrm{Ch}}=\alpha^{2}m_{e}^{2}c^{4}\left(1+ 1/\alpha \chi_{e}^{2}\right) /\hbar $ for the Cherenkov and $ P_{\mathrm{C}}=\sqrt{3}\alpha m_{e}^{2}c^{4}F\left(\chi_{e},\chi_{\gamma}\right) /2\pi\hbar  $ for the Compton radiation, where $ F\left(\chi_{e},\chi_{\gamma}\right) $ is the quantum-corrected synchrotron spectrum function \cite{Ritus1985}. For the above-mentioned parameters we obtain $ P_{\mathrm{Ch}}\approx0.15P_{\mathrm{C}} $. The average formation time for pair production by such a photon is on the order of $ \sim4\hbar E_{\mathrm{S}} /\alpha m_{e}c^{2} E_{0} $ i.e. $\sim 10^{-2}T $ \cite{Ritus1985}. Provided that these Cherenkov photons create pairs at the same rate as the Compton ones, the total number of positrons with the average initial energy $ \mathcal{E}_{e}^{\mathrm{c}}/24 \approx20~\mathrm{MeV} $ (assuming pair production rate has a maximum at producing electron and positron with equal energy for this value of $\chi_{\gamma}$) should be higher by 15\% compared to the case when only Compton radiation is considered.
These positrons should be created within the cone of opening angle $ \theta_{\mathrm{Ch}}\approx2\sqrt{4\alpha \left( E_{0}/E_{\mathrm{S}}\right) ^{2}+\left( m_{e}c^{2}/\mathcal{E}_{e}^{\mathrm{c}}\right) ^{2}}\approx3\times10^{-3} $, see Fig.~\ref{fig:Fig_05} \cite{Ritus1985}. 
%
%
This might serve as an indication of the vacuum polarization effect presence.

In conclusion, we estimate the average electron energy in the center of the counter-propagating intense EM wave in the limit $ \chi_{e}\gg1 $ as well as the condition for electron reflection.
%
%
%
The results show that reaching the center of the multi-PW laser pulse is 
attainable with 10-100s GeV electron even when radiation reaction is considered.
The roles of laser pulse duration and laser wavelength on reaching the laser 
pulse amplitude are quantitatively assessed and are in good agreement with the 
simulations.
The conditions on electron initial energy for reaching  the center of the laser 
pulse and emitting the Cherenkov photons are provided with respect to the laser 
pulse intensity and duration.
%
%

Portions of this research were carried out at ELI Beamlines, a European user 
facility operated by the Institute of Physics of the Academy of Sciences of the 
Czech Republic.
This work is supported by the project High Field Initiative (HIFI) CZ.02.1.01/0.0/0.0/15\_003/0000449 from European
Regional Development Fund (ERDF) and Czech Science Foundation project No. 18-09560S. SSB acknowledges support from the U.S. Department of Energy Office of Science Offices of High Energy Physics and Fusion Energy Sciences (through LaserNetUS), under Contract No. DE-AC02-05CH11231. The simulations were performed at the clusters ECLIPSE at ELI Beamlines and IT4Innovations.

\bibliography{electron_laser}{}
\bibliographystyle{apsrev4-2}
\end{document}